\documentclass[preprint,prd,tightenlines,superscriptaddress]{revtex4-1}

\usepackage{graphicx} 
\usepackage{dcolumn}  
\usepackage{colordvi}
\usepackage{color}
\usepackage{epstopdf}
\usepackage{amssymb}
\usepackage{url}
\graphicspath{{ps}}
\usepackage{hyperref}
\usepackage{tabularx}
\usepackage{float} 
\usepackage{bm}
\usepackage{subfigure} 
\usepackage{hyperref}
\usepackage{tasks}

\usepackage{placeins}

\usepackage[english]{babel}

\begin{document}

\clubpenalty = 10000  
\widowpenalty = 10000 

\vspace*{-3\baselineskip}
\resizebox{!}{3cm}{\includegraphics{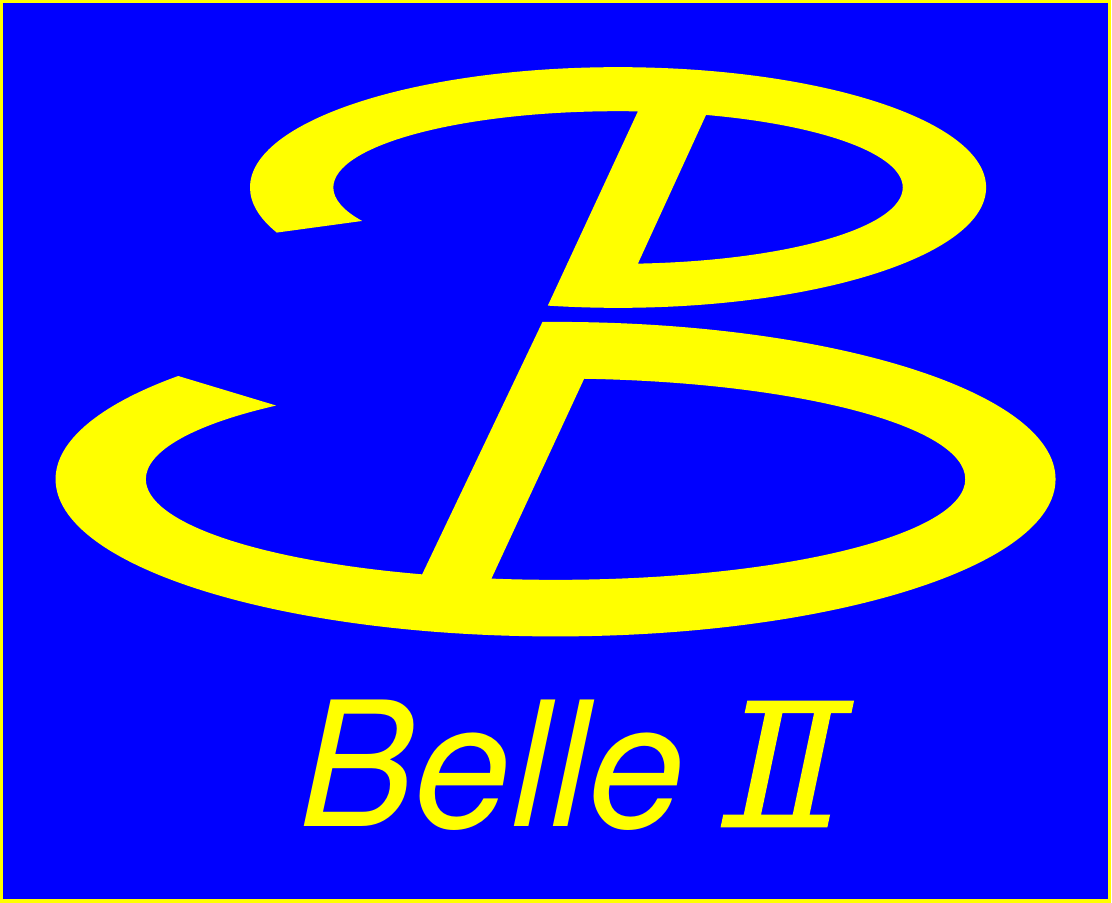}}

\vspace*{-5\baselineskip}
\begin{flushright}
BELLE2-CONF-2022-006

\today
\end{flushright}

\quad\\[0.5cm]

\title { Measurements of the branching fraction, isospin asymmetry, and lepton-universality ratio in {\boldmath $B \to J/\psi K$} decays at Belle II}


  \author{F.~Abudin{\'e}n}
  \author{I.~Adachi}
  \author{R.~Adak}
  \author{K.~Adamczyk}
  \author{L.~Aggarwal}
  \author{P.~Ahlburg}
  \author{H.~Ahmed}
  \author{J.~K.~Ahn}
  \author{H.~Aihara}
  \author{N.~Akopov}
% ...

\collaboration{Belle II Collaboration}

\begin{abstract}
We report a study of $B \to J/\psi(\ell^{+}\ell^{-}) K$ decays, where $\ell$ represents an electron or a muon, using $e^{+}e^{-}$ collisions at the $\Upsilon(4S)$ resonance.
The data were collected by the Belle~II experiment at the SuperKEKB asymmetric-energy collider during 2019--2021, corresponding to an integrated luminosity of $189$\,fb$^{-1}$. 
The measured quantities are the branching fractions (${\mathcal B}$) of the decay channels $B^{+} \to J/\psi(e^{+}e^{-}) K^{+}$, $B^{+} \to J/\psi(\mu^{+}\mu^{-}) K^{+}$, $B^{0} \to J/\psi(e^{+}e^{-}) K^{0}_{S}$, and $B^{0} \to J/\psi(\mu^{+}\mu^{-}) K^{0}_{S}$; the
lepton-flavor-dependent isospin asymmetries for the electron [$A_{I}\left(B \to J/\psi(e^{+}e^{-}) K\right)$] and muon [$A_{I}\left(B \to J/\psi(\mu^{+} \mu^{-}) K\right)$]
channels; and the ratios of branching fractions between the  muon and electron  channels for the 
charged [$R_{K^{+}}\left(J/\psi\right)$] and neutral kaon [$R_{K^{0}}\left(J/\psi\right)$] case.
We obtain
\begin{eqnarray*}
\mathcal{B} \left( B^{+} \to J/\psi(e^{+} e^{-}) K^{+}\right) &=& (6.00 \pm 0.10 \pm 0.19) \times 10^{-5},\\
\mathcal{B} \left( B^{+} \to J/\psi(\mu^{+} \mu^{-}) K^{+}\right) &=& (6.06 \pm 0.09 \pm 0.19) \times 10^{-5},\\
\mathcal{B} \left( B^{0} \to J/\psi(e^{+} e^{-}) K_{S}^{0} \right) &=& (2.67 \pm 0.08 \pm 0.12) \times 10^{-5},\\
\mathcal{B} \left( B^{0} \to J/\psi(\mu^{+} \mu^{-}) K_{S}^{0} \right) &=& (2.78 \pm 0.08 \pm 0.12) \times 10^{-5},\\
A_{I} \left( B \to J/\psi(e^{+} e^{-}) K\right) &=& -0.022 \pm 0.016 \pm 0.030,\\
A_{I} \left( B \to J/\psi(\mu^{+} \mu^{-}) K\right) &=& -0.006 \pm 0.015 \pm 0.030,\\
R_{K^{+}}\left(J/\psi\right) &=& 1.009 \pm 0.022 \pm 0.008, \text{ and}\\
R_{K^{0}}\left(J/\psi\right) &=& 1.042 \pm 0.042 \pm 0.008,
\end{eqnarray*}
where the first uncertainties are statistical and the second are systematic. The measurements are consistent with the world-average values. 

\end{abstract}


\maketitle

{\renewcommand{\thefootnote}{\fnsymbol{footnote}}}
\setcounter{footnote}{0}

\pagebreak

\section{Introduction}

The decays $B \to K \ell^+ \ell^-$, where $\ell$ stands for an electron or a muon, are flavor-changing-neutral-current processes.
Being governed by a $b \to s$ quark-level transition, these decays are forbidden at tree level in the standard model (SM) of particle physics~\cite{GIM}, but can proceed through $b\to s\ell^+ \ell^-$ loop  amplitudes at lowest order. Various SM extensions~\cite{BSM1, BSM2} predict new particles that contribute to the processes, altering the values of observables from their SM predictions. These possibilities make $B \to K \ell^+ \ell^-$ decays a sensitive probe for beyond-the-SM physics. 

One of the key predictions of the SM is that the coupling strengths of electroweak gauge bosons to charged leptons $e$, $\mu$, and $\tau$ are the same, a property known as lepton-flavor universality (LFU). Accordingly, the ratio of branching fractions of $B \to K \mu^+ \mu^-$ to $B \to Ke^+ e^-$ decays, called $R_K$, is expected to be close to unity~\cite{rkeq1, rkerr}. A recent measurement of $R_{K}$ by the LHCb Collaboration reported a $3.1\sigma$ discrepancy with respect to its SM prediction in the range of dilepton mass squared $q^2 \in \left( 1.1, 6.0\right)$\,GeV$^2$/$c^4$~\cite{lhcb}. On the other hand, the measurements reported by the Belle experiment are consistent with both the SM and the LHCb result~\cite{Belle_Rk}, albeit with significantly less precision compared to the latter.

In this report, we describe measurements of branching fractions $\mathcal{B}\left(B\to J/\psi K \right)$,  LFU ratios
\begin{equation}
    \label{eqn:rk}
    R_{K}{\left(J/\psi\right)} = 
    \frac{\mathcal{B} \left(B\to J/\psi(\mu^{+}\mu^{-})K\right)}{\mathcal{B} \left(B\to J/\psi(e^
    {+}e^{-})K\right)},
\end{equation}
 and  isospin asymmetries \begin{equation}
    \label{eqn:AI}
    A_{I} = 
    \frac{\Gamma[B^{0}\to J/\psi(\ell^{+}\ell^{-})K^{0}] - \Gamma[B^{+}\to J/\psi(\ell^{+}\ell^{-})K^{+}]}{\Gamma[B^{0}\to J/\psi(\ell^{+}\ell^{-})K^{0}] + \Gamma[B^{+}\to J/\psi(\ell^{+}\ell^{-})K^{+}]},
\end{equation}
performed using data recorded by the Belle~II experiment. In contrast to suppressed, charmless $B\to K\ell^+ \ell^-$ decays, the $B\to J/\psi(\ell^+ \ell^-)K$ decays involve a favored $b\to c$ tree-level transition. Hence, contributions from beyond-the-SM physics are expected to have a negligible impact. Since the branching fraction of $B\to J/\psi(\ell^+ \ell^-)K$ is two orders of magnitude larger than that of $B\to K\ell^+ \ell^-$ decays and both channels share the same final-state particles, the former decays constitute an excellent control sample for studies of the latter. Therefore, the measurement of $R_{K}\left(J/\psi\right)$ and its consistency with unity would be a strong validation of the future $R_K$ measurement in the charmless counterpart of $B\to K\ell\ell$ decays. Throughout the report, charge conjugate processes are implicitly included.

Section~\ref{b2detector} gives a brief introduction to the Belle II detector and the data samples used in the analysis. Sections~\ref{evt_select} and \ref{bkg} describe the event selection and background suppression methods. Section~\ref{fit} explains the signal yield determination and Section~\ref{measure_obs} presents the measurements of observables. Section~\ref{syst} discusses systematic uncertainties and a summary is provided in Section~\ref{summary}.

\section{The Belle II Detector and Data Sample}     
\label{b2detector}
The Belle II detector is a large-solid-angle magnetic spectrometer designed to study final-state particles produced in energy-asymmetric $e^+e^-$ collisions delivered by SuperKEKB~\cite{SuperKEKB} at a center-of-mass energy corresponding to the mass of the $\Upsilon(4S)$ resonance.  The detector is composed of several subdetectors arranged in a cylindrical structure around the beam pipe. The subdetectors are two layers of silicon pixel detectors (PXD), four layers of double-sided silicon strip detectors (SVD), a 56-layer central drift chamber (CDC), a time-of-propagation counter (TOP) in the barrel region, a proximity focusing ring imaging Cherenkov counter (ARICH) in the forward region, and an electromagnetic calorimeter (ECL) comprising CsI(Tl) crystals. All these subdetectors are located inside a superconducting solenoid that provides a 1.5\,T magnetic field. The return yoke of the magnet is instrumented with plastic scintillators and resistive plate chambers to identify $K^{0}_{L}$ mesons and muons (KLM). Further details about the detector can be found in Ref.~\cite{b2tdr}. 

The data sample used in this analysis amounts to an integrated luminosity of 189\,fb$^{-1}$, which is equivalent to $198\times10^6$ $B \overline{B}$ events.
To study the properties of signal events, to optimize selection criteria, and to determine the detection efficiency, $2 \times 10^6$ simulated signal events are generated for each of the $B^{+} \to J/\psi(e^{+}e^{-}) K^{+}$, $B^{+} \to J/\psi(\mu^{+} \mu^{-}) K^{+}$, $B^{0} \to J/\psi(e^{+}e^{-})  K_{S}^{0}$ and $B^{0} \to J/\psi(\mu^{+} \mu^{-}) K_{S}^{0}$ decay channels.
 In addition, we use a simulated sample equivalent to 1\,ab$^{-1}$ of inclusive $B\overline{B}$ and $q\overline{q}$ continuum events to study background processes. Here, $q$ denotes a $u$, $d$, $s$, or $c$ quark.  The $B$ meson decays are generated using the \textsc{EvtGen}~\cite{EVTGEN} package, where the final-state-radiation effects from charged particles are incorporated using the \textsc{PHOTOS} package~\cite{PHOTOS}. Continuum background events are simulated  using the \textsc{KKMC}~\cite{KKMC} generator interfaced with \textsc{PYTHIA}~\cite{PYTHIA}, and \textsc{Geant4}~\cite{GEANT4} simulates the interaction of generated particles with the detector. The Belle II analysis software framework~\cite{BASF2} is used to process both data and simulation events.

\section{Event Selection and  Reconstruction}
\label{evt_select}
The reconstruction follows a hierarchical procedure, where we first reconstruct neutral and charged particles using information from ECL and tracking subdetectors. Information from all subdetectors except the PXD and SVD is used to identify charged particles as kaons, electrons, and muons. Next, we take pairs of oppositely-charged leptons to reconstruct $J/\psi$ candidates.  Pairs of charged particles are used to reconstruct $K^{0}_{S}$ candidates. Finally, we combine  $J/\psi$ candidates with  kaons, which are either $K^{+}$ or  $K^{0}_{S}$ candidates, to reconstruct $B$ mesons. The remainder of this section describes each step in detail.

The information from PXD, SVD, and CDC is used to reconstruct charged particles. We require that the distance of closest approach to the $e^{+} e^{-}$ interaction point in the plane transverse to the $z$ axis be less than 2.0\,cm and that
along the $z$ axis be less than 4.0\,cm, to select charged particles originating from a region around the collision point.
The $z$ axis is collinear with the symmetry axis of the solenoid approximately in the $e^{-}$ beam direction.
To suppress contamination from low-multiplicity $e^{+}e^{-}\to e^{+}e^{-}\ell^{+}\ell^{-}$, $e^{+}e^{-}\to e^{+}e^{-}\gamma$ and $e^{+}e^{-}\to \mu^{+}\mu^{-}\gamma$  backgrounds, we require that the event contains at least five charged particles, each with a transverse momentum greater than 100\,MeV/$c$, and that the normalized second Fox--Wolfram moment of the event is less than 0.7~\cite{R2}.

From the set of selected tracks, we distinguish charged kaons from pions using specific ionization information from the CDC, arrival time of the two-dimensional information of a Cherenkov ring image from the TOP, and the number of detected Cherenkov photons from the ARICH. The  kaon identification efficiency is 86\% with a pion misidentification rate of 7\%, estimated for kaons within a momentum range $(0.5,4.5)\:\text{GeV}/c$ and polar-angle acceptance $(0.49, 2.62)$\,rad. The penetration depth and transverse scattering pattern in the KLM are used to identify muon candidates. Furthermore, we require a minimum momentum of 0.8~GeV/$c$ to ensure that the candidate reaches the KLM. These criteria result in a muon identification efficiency of 87\% with a pion misidentification rate of 7\%, calculated using muons within the momentum range $(0.4,6.5)\:\text{GeV}/c$ and polar-angle acceptance $(0.40, 2.60)$\,rad. Electron candidates are identified mainly using the ratio of the ECL energy to the momentum, the ECL shower shape, and the position matching of the track with the calorimetric cluster. To ensure that electrons reach the ECL, a momentum threshold of 0.5~GeV/$c$ is applied. The electron identification efficiency is 94\% and the pion misidentification rate is around 2\%, estimated for electrons within the momentum range $(0.4,7.0)\:\text{GeV}/c$ and polar-angle acceptance $(0.22, 2.71)$\,rad. Relativistic electrons tend to lose their energy by radiating photons via the bremsstrahlung process. For a given electron depositing energy in the ECL, the four-momenta of all photons detected within a cone of 50\,mrad with respect to the initial momentum direction of the electron and having an energy greater than 50\,MeV are added to that of the electron to recover bremsstrahlung energy loss.

The $K^{0}_{S}$ candidates are reconstructed from two oppositely charged particles, assumed to be pions. A kinematic fit is performed on the $K^{0}_{S}$ vertex to ensure that the pion tracks originate from a common vertex. We retain candidates with an invariant mass in the range $(0.487,0.508)\,{\rm GeV}\!/c^2$, which is $\pm 3 \sigma$ resolution around the known $K^{0}_{S}$ mass. In addition, momentum-dependent selections are applied on the $K^{0}_{S}$ flight length in the transverse plane, the azimuthal angle between the momentum vector and the vector connecting the interaction point with the decay vertex of the $K^{0}_{S}$ candidate, and the difference between the distances of closest approach of the two pion tracks along the $z$ axis. 

We reconstruct $J/\psi$ candidates by combining two oppositely-charged leptons of the same flavor.
The invariant mass of the $J/\psi$ candidate must lie in the range $(2.91,3.19)\,{\rm GeV}\!/c^2$ and $(2.96,3.19)\,{\rm GeV}\!/c^2$ for the electron and muon channel, respectively. Imperfect recovery of the energy lost due to bremsstrahlung motivates a wider asymmetric interval around the known $J/\psi$ mass in the former case.

We reconstruct $B$ candidates by combining a kaon candidate ($K^{\pm}$ or $K^{0}_{S}$) with a $J/\psi$ candidate. A vertex fit is performed on the $B$ candidates. We require the fit to converge in order to suppress background coming from a random combination of charged particles. To distinguish signal from background, the following two kinematic variables are used:
\begin{equation}
 M_{\rm bc}   = \sqrt{s/4-\vec{p}^{\,\ast2}_{B}} \text{   and}\\
\end{equation}
\begin{equation}
\Delta E  = E^{\ast}_{B}-\sqrt{s}/2,
\end{equation}
where ($E_{B}^{\ast} , \vec{p}^{\,\ast}_{B}$) is the four-momentum of the $B$ candidate in the center-of-mass frame and $\sqrt{s}$ is the center-of-mass energy. For a correctly reconstructed $B$ candidate, $M_{\rm bc}$ and $\Delta E$ distributions are expected to peak around the known $B$ mass and zero, respectively.  We retain candidate events satisfying  $M_{\rm bc} \in \left(5.20,5.29\right)$\,GeV/$c^{2}$ and $|\Delta E| < 0.20$\,GeV.

\section{Background suppression}
\label{bkg}
Following the selection, we find a major background contribution coming from $B\to  J/\psi K^{*}(K\pi)$ decays, in which $B$ candidates are reconstructed without the final-state pion.
As a result, the $\Delta E$ distribution for such events peaks  below $-0.10\,\mathrm{GeV}$. A selection of $\Delta E>-0.10\,\mathrm{GeV}$ thus rejects about 98\% of $B\to  J/\psi K^{*}(K \pi)$ background events.

The background for the charged $B$ channels is mostly from $B^{+} \to  J/\psi \pi^{+}$ decays where the pion is misidentified as a kaon.  These decays  peak inside the $B$ meson mass region of the $M_{\rm bc}$ distribution. Unlike $M_{\rm bc}$, the $\Delta E$ variable is directly sensitive to the mass hypotheses of the $B$ candidate decay products. Hence, these events show a shifted peak in the $\Delta E$ distribution with respect to signal events. The other background contribution arises from a random combination of $B$ decay products, which does not peak in either the $M_{\rm bc}$ or $\Delta E$ distribution.

After applying all the selection criteria, the average candidate multiplicity ranges between 1.005 and 1.010 depending on the decay channel. In case of multiple candidates, we retain the one with the highest $\chi^{2}$ probability of the $B$ vertex fit. The efficiency obtained from simulation
to select the correctly reconstructed signal from an event with multiple reconstructed $B$ candidates varies from 78 to 83\% depending on the decay channel.

\section{Signal Extraction}
\label{fit}
The signal yield is obtained from an extended maximum-likelihood fit to the  $M_{\rm bc}$ and $\Delta E$  distributions. 
For a data sample containing $N$ candidates, the likelihood function is given as
\begin{equation}
{\mathcal L}(\vec{\alpha},\vec{\beta}, \vec{n}) = \frac{e^{-\sum^{}_{j} n_{j}}}{N!} \prod^{N}_{i=1}   \sum^{}_{j} n_{j}\mathcal{P}_{j}(M_{\rm bc}^{i};{\vec{\alpha}_{j}})\mathcal{Q}_{j}(\Delta E^{i};{\vec{\beta}_{j}}),    
\end{equation}

where $\mathcal{P}_{j}$ and $\mathcal{Q}_{j}$ are the probability density functions (PDFs) of the $M_{\rm bc}$ and  $\Delta E$  distributions, and $n_{j}$ is the number of events corresponding  to the $j\text{th}$ component. We assume the $M_{\rm bc}$ and  $\Delta E$  distributions to be uncorrelated. The fit performed for the neutral $B$ channels employs two components, one for the correctly reconstructed signal and the other for misreconstructed background events. The fit to charged channels has an additional  component to model $B^{+}\to J/\psi \pi^{+}$ background events. 
The argument $M_{\rm bc}^{i}$ and  $\Delta E^{i}$ denote the  $M_{\rm bc}$ and $\Delta E$  values for the $i\text{th}$ candidate;
$\vec{\alpha}_{j}$ and $\vec{\beta}_{j}$ are  the set of shape parameters for $\mathcal{P}_{j}$ and $\mathcal{Q}_{j}$, respectively. 
The likelihood function is maximized with respect to $\vec{\alpha}_{j}$, $\vec{\beta}_{j}$, and $n_{j}$. 

\begin{figure}[!ht]
	\centering
	\subfigure
	{
		\includegraphics[scale=0.39]{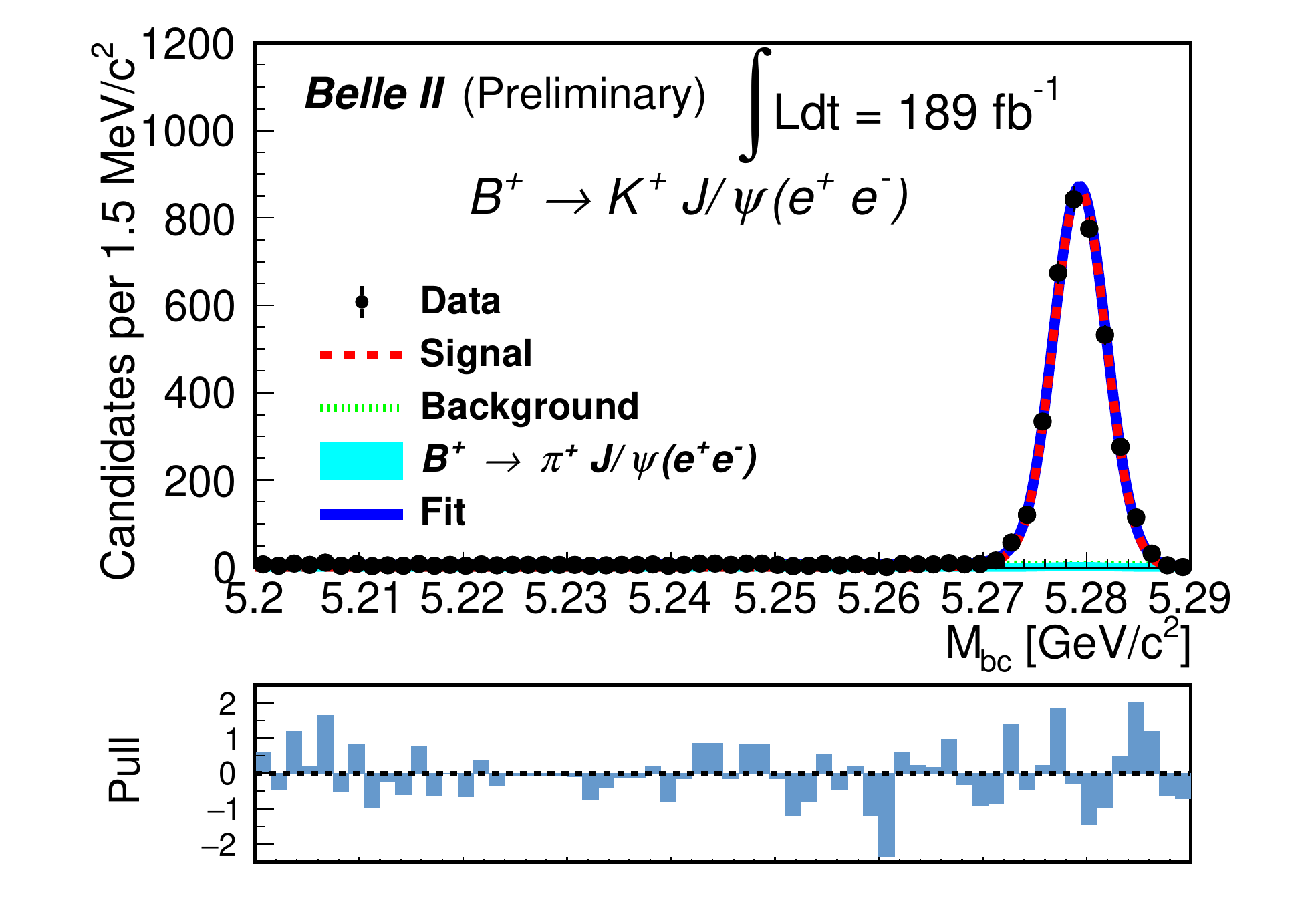}
	}%
	\subfigure
	{   
		\includegraphics[scale=0.39]{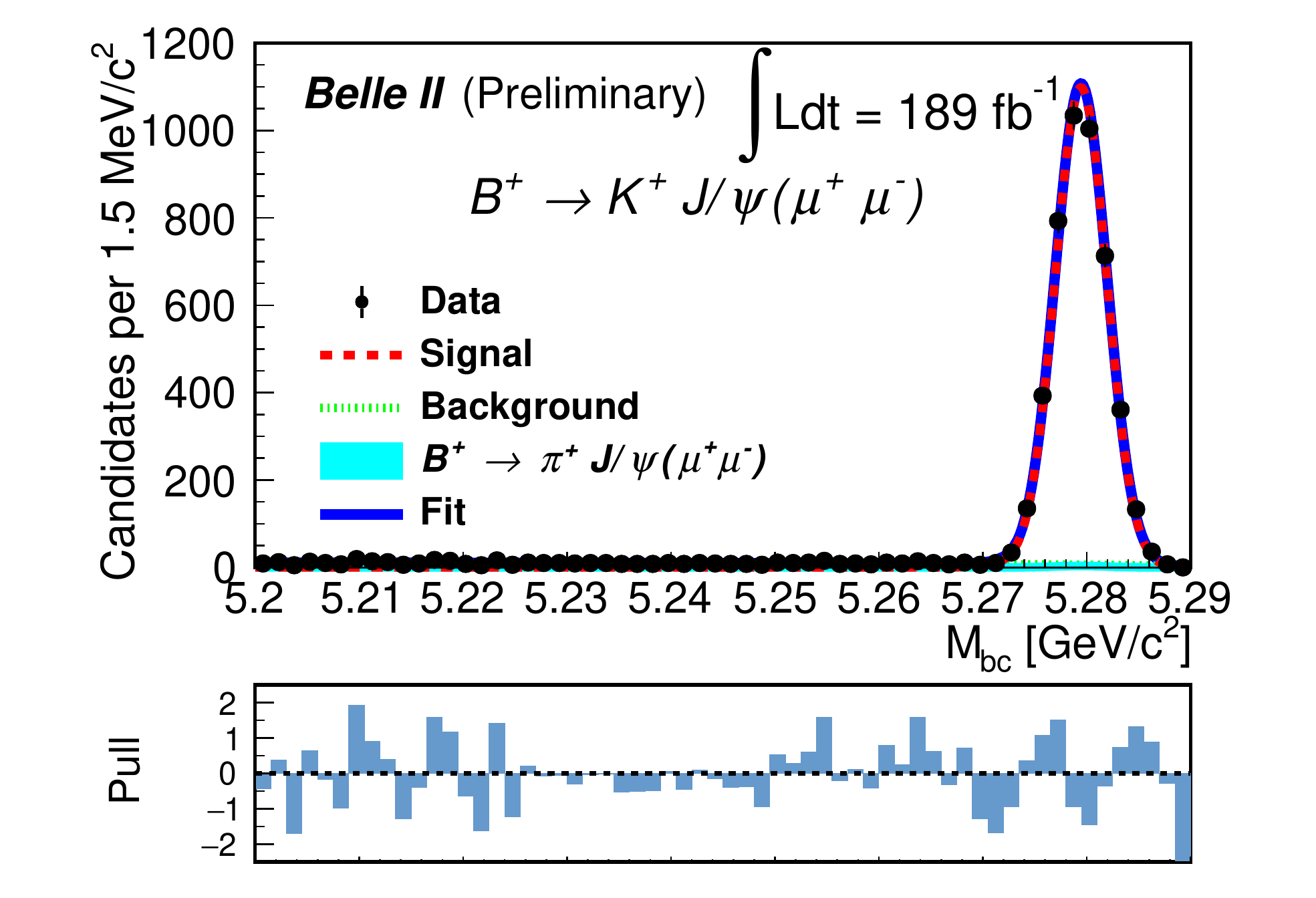}
	}\\
	\subfigure
	{   
		\includegraphics[scale=0.39]{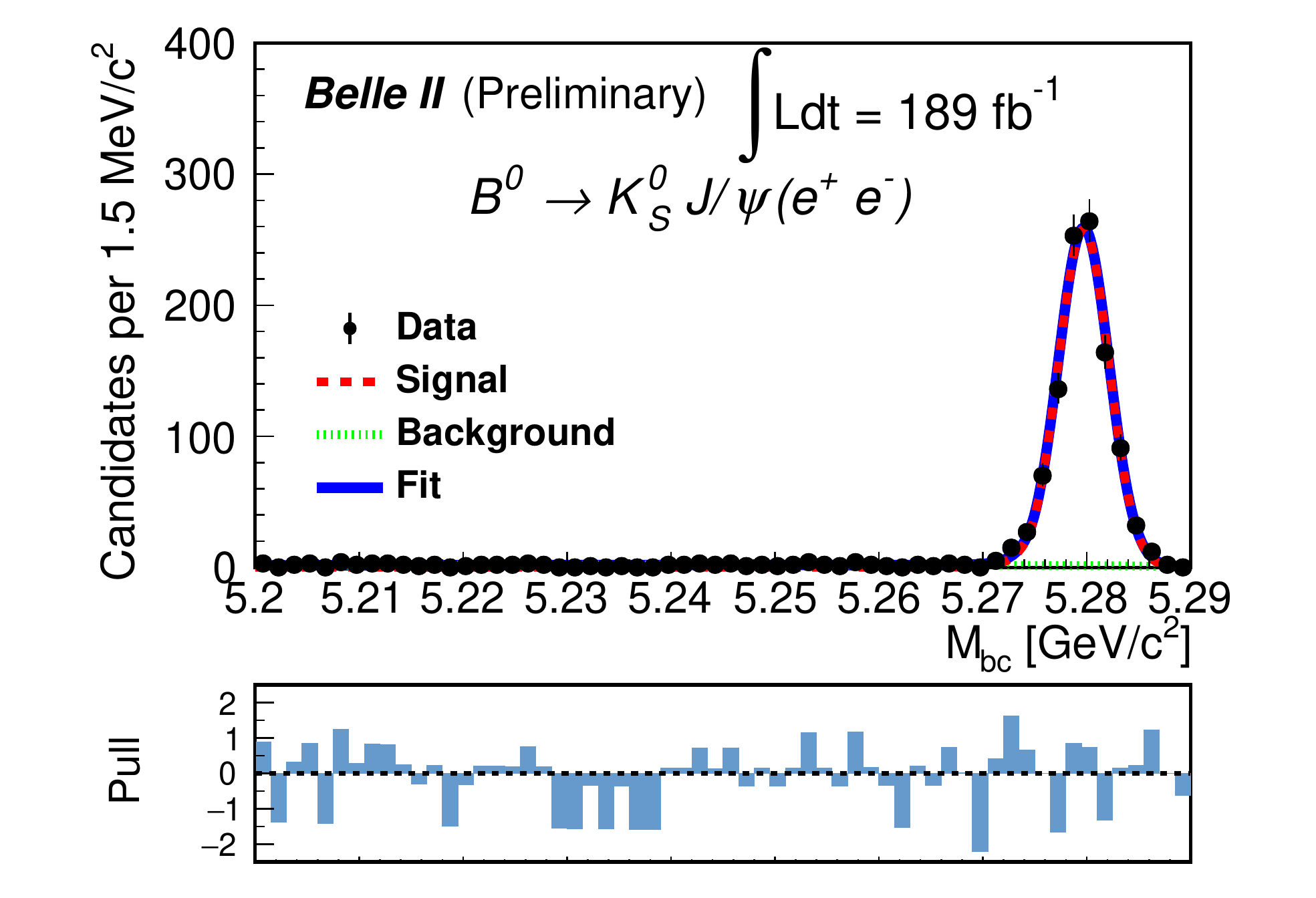}
	}%
	\subfigure
	{   
		\includegraphics[scale=0.39]{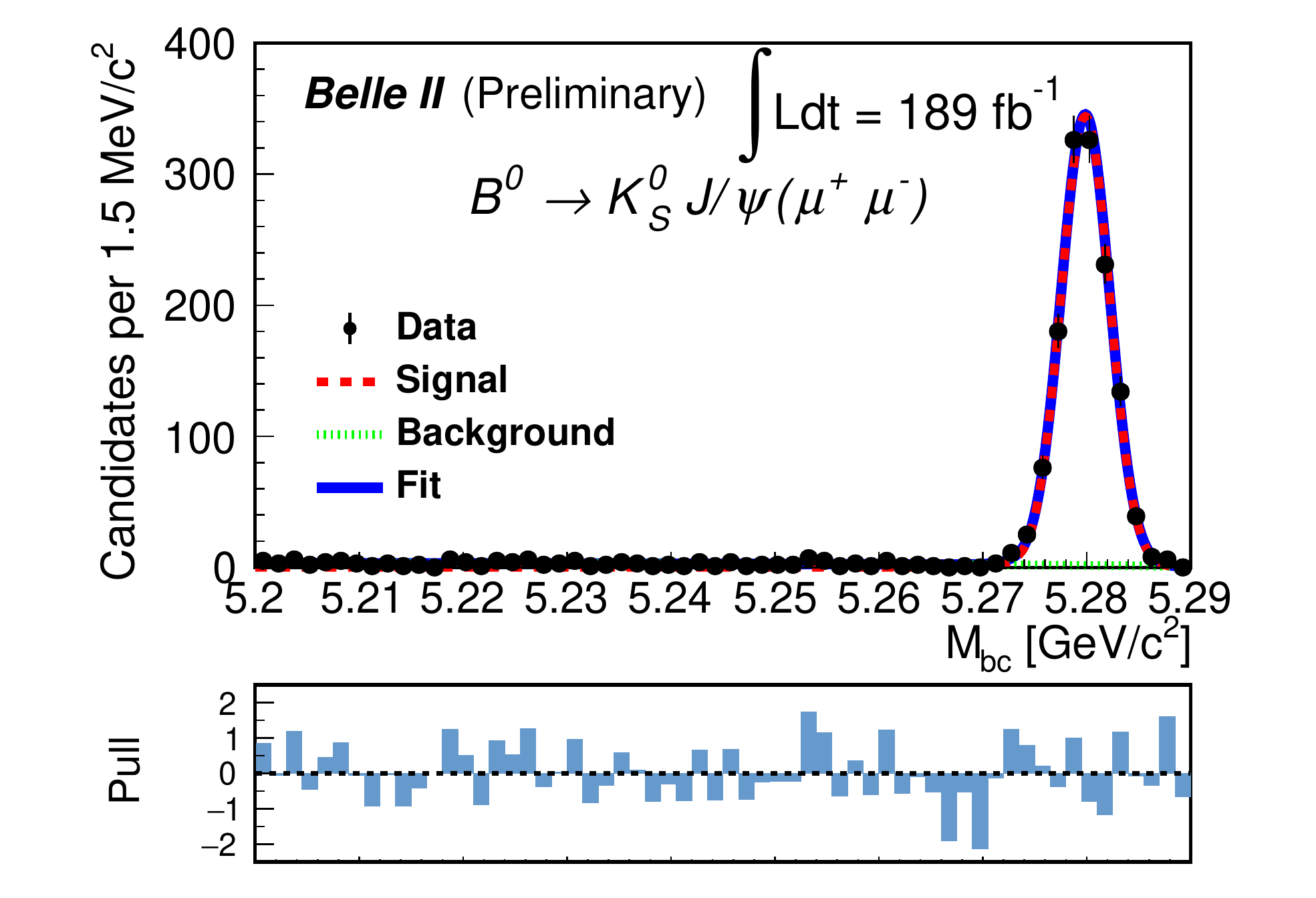}
	}

	\caption{$M_{\rm bc}$ distributions for each $B \to J/\psi(\ell^+ \ell^-)K$ channel with the fit result superimposed (top) and pull distribution with respect to the fit result (bottom), where the pull is defined as the difference between the fit result and the value of the distribution in a bin, divided by the estimated uncertainty in that bin. Black dots with error bars denote the data, blue curves denote the total fit, dashed red curves are the signal component, dotted green curves are the background component, and filled cyan regions in the charged channels are the $B^+\to J/\psi\pi^+$ component.}
	\label{mbc fit}
\end{figure}

\begin{figure}[!ht]
	\centering
	\subfigure
	{
		\includegraphics[scale=0.39]{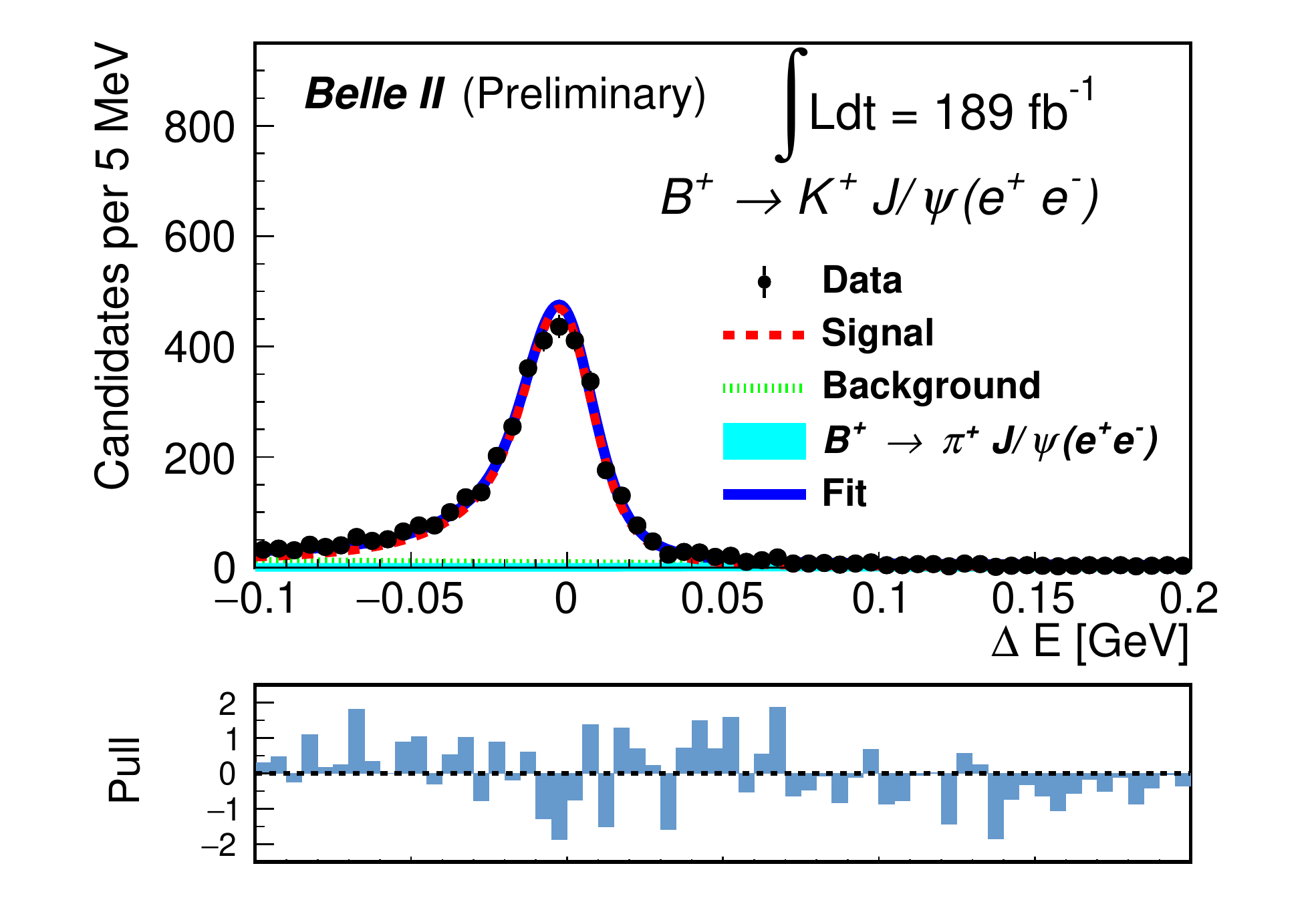}
	}%
	\subfigure
	{   
		\includegraphics[scale=0.39]{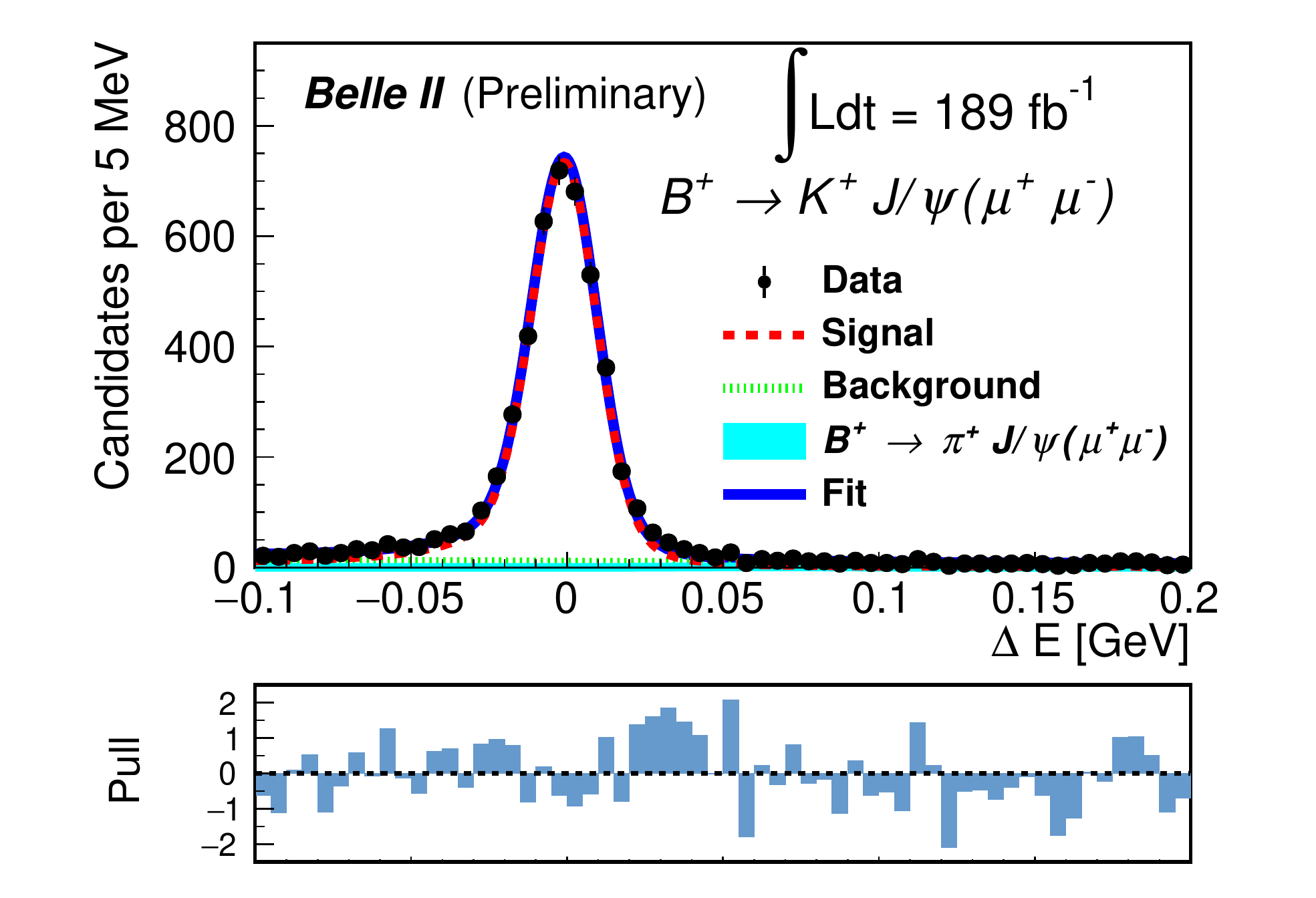}
	}\\
	\subfigure
	{   
		\includegraphics[scale=0.39]{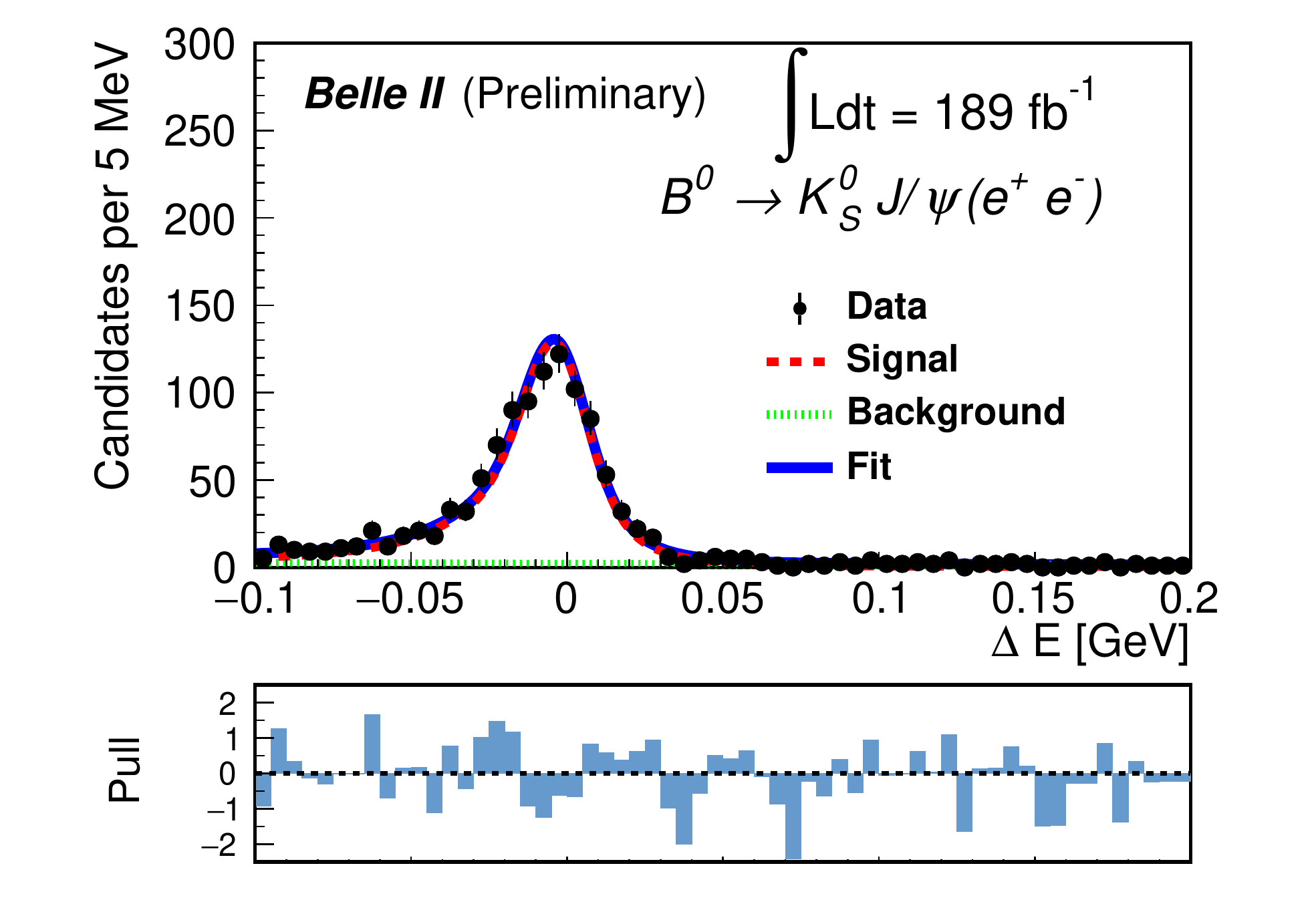}
	}%
	\subfigure
	{   
		\includegraphics[scale=0.39]{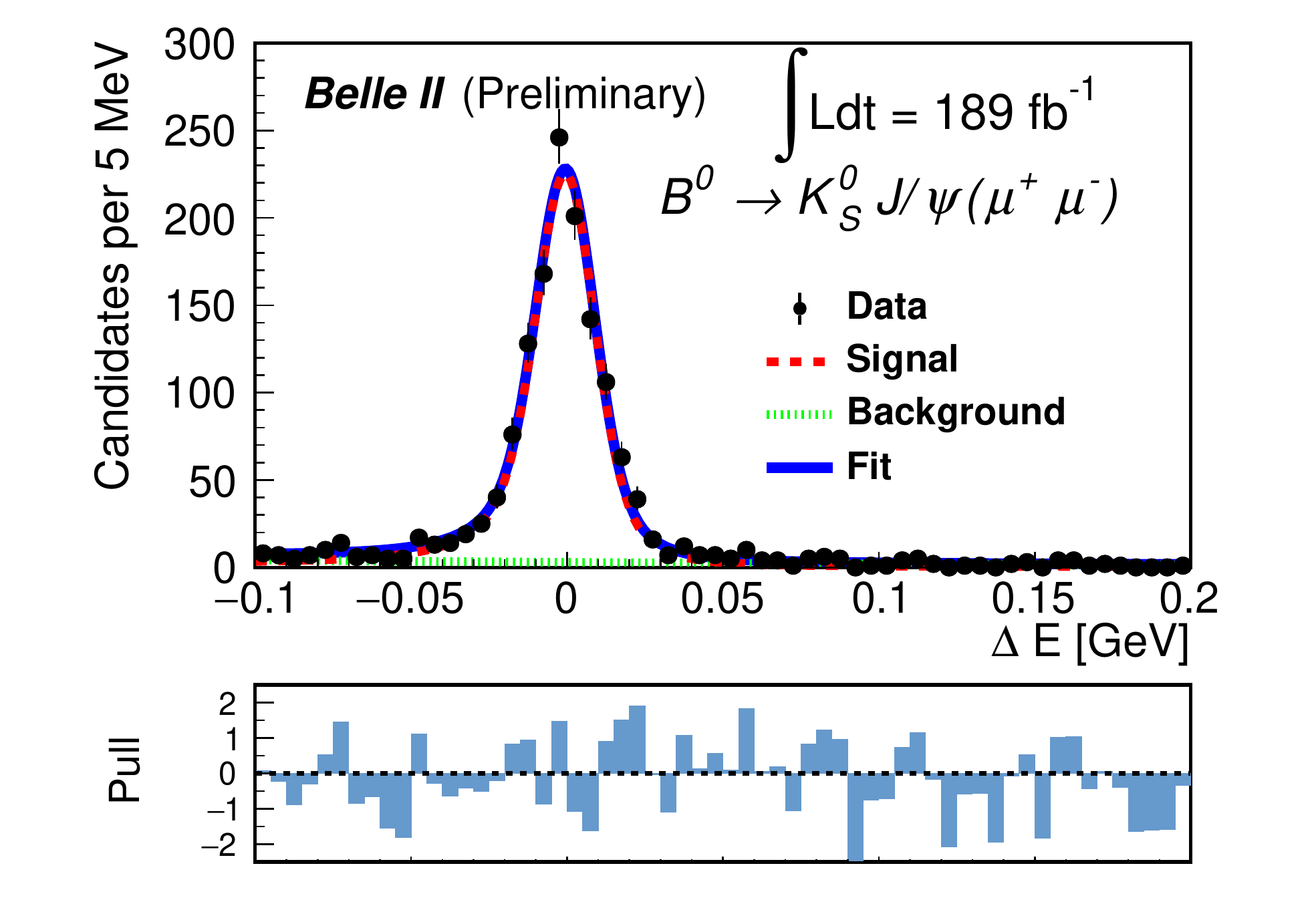}
	}

	\caption{$\Delta E$ distributions for each $B \to  J/\psi(\ell^+ \ell^-) K$ channel with the fit result superimposed (top) and pull distribution with respect to the fit result (bottom). Black dots with error bars denote the data, blue curves denote the total fit, dashed red curves are the signal component, dotted green curves are the background component, and filled cyan regions in the charged channels are the $B^+ \to J/\psi \pi^+ $ component.}
	\label{de fit}
\end{figure}

The PDF that describes the $\Delta E$ distribution of the signal component is determined using correctly reconstructed events from the simulated signal sample. The signal PDF is a sum of an empirical function~\cite{RooCruijiff} and a Gaussian. The signal $M_{\rm bc}$ PDF is parametrized by a function introduced by the Crystal Ball Collaboration~\cite{CB}.
This function accounts for radiative tails in the distribution as well as for the finite photon energy resolution. Other than the mean and width, the remaining signal shape parameters are fixed to those obtained from the simulated sample.
The $\Delta E$  distribution of  background is modeled by an exponential function, whose shape parameter is determined from the fit to data. The  background $M_{\rm bc}$ distribution is parameterized by a threshold function introduced by the ARGUS Collaboration~\cite{ARGUS}. The kinematic endpoint of the ARGUS PDF is fixed to $\sqrt{s}/2 \approx 5.291$~GeV$/c^{2}$ with the other shape parameter determined from the fit to data.
The $M_{\rm bc}$  and $\Delta E$ distributions for the $B^{+} \to J/\psi \pi^{+} $ component are modeled using a Gaussian function, whose shape parameters are also fixed to those obtained from the simulated sample. The yield of $B^{+} \to  J/\psi \pi^{+}$ events is fixed to a value estimated using the known branching fraction~\cite{PDG} and the $\pi \to K$ misidentification rate obtained in auxiliary data. We estimate around 11 and 14 events coming from $B^{+} \to J/\psi(e^{+}e^{-})\pi^{+}$ and $B^{+} \to J/\psi(\mu^{+}\mu^{-})\pi^{+}$ background, respectively. The distributions of $M_{\rm bc}$ and $\Delta E$ for each $B$ channel are shown in Figs.~\ref{mbc fit} and \ref{de fit}, respectively; the fit results are superimposed.

To check for potential fit biases, an ensemble of 1000 simulated data samples is prepared from 1~ab$^{-1}$ of fully-simulated events using sampling with replacement~\cite{bootstrap}. The number of signal and background events in each data sample are drawn from a Poisson distribution, with the mean corresponding to the total number of events expected in the data. Each of these simulated data samples is fit. The estimators are unbiased and have Gaussian distributions, validating our assumption that in simulated data $M_{\rm bc}$ and  $\Delta E$  distributions are uncorrelated.

\section{Measurement of observables}
\label{measure_obs}
The branching  fractions are determined using the relation
 
 \begin{equation}
\label{eqn:relatn0}
 \mathcal{B} = \frac{n_{\rm sig}}{2\, N_{B\overline{B}}\, f^{i} \, \epsilon },
 \end{equation}
 where  ${n}_{\rm sig}$ is the signal yield determined by the fit, $N_{B\overline{B}}$ is the number of $B\overline{B}$ events,
  $\epsilon$ is the signal selection efficiency, $f^{i}$ is  $f^{\pm}$ for the charged channels and $f^{00}$ for the neutral channels; $f^{\pm} (f^{00})$ is the branching fraction of $\Upsilon(4S)$ to charged (neutral) $B \overline{B}$ pairs assuming $\mathcal{B} \left(\Upsilon(4S) \to B\overline{B}\right) = 1$. We use the values  $f^{00}=0.487 \pm 0.013$ and $f^{\pm} = 1- f^{00} = 0.513 \pm    0.013$ reported in Ref.~\cite{fpmf00} rather than the world average~\cite{HFLAV}, as the latter is dominated by measurements that assume isospin symmetry in $B \to J/\psi K$ decays.
  
 We calculate $R_K$ using the relation
 \begin{equation}
\label{eqn:rkjpsi}
    R_{K}{\left(J/\psi\right)} = 
    \frac{\mathcal{B}\left(J/\psi(\mu^{+}\mu^{-})K\right)}{\mathcal{B}\left(J/\psi(e^
    {+}e^{-})K\right)}=
    \frac{n_{\rm sig}^{J/\psi(\mu^{+}\mu^{-})K}}{n_{\rm sig}^{J/\psi(e^{+}e^{-})K}} \times
    \frac{\epsilon^{J/\psi(e^{+}e^{-})K}}{\epsilon^{J/\psi(\mu^{+}\mu
    ^{-})K}},
\end{equation}
    and $A_I$ using the relation
 
\begin{equation}
\label{eqn:ai}
    A_{I} = \frac{2 \, (\tau_{B^{+}}/\tau_{B^{0}}) (f^{\pm}/f^{00}) (n_{\rm sig}/\epsilon)|_{ J/\psi(\ell^+ \ell^-)K_{S}^{0}} -(n_{\rm sig}/\epsilon)|_{ J/\psi (\ell^+ \ell^-)K^{+}}}{2  \, (\tau_{B^{+}}/\tau_{B^{0}}) (f^{\pm}/f^{00}) (n_{\rm sig}/\epsilon)|_{ J/\psi(\ell^+ \ell^-)K_{S}^{0}} +(n_{\rm sig}/\epsilon)|_{ J/\psi (\ell^+ \ell^-)K^{+}}},
\end{equation}
where $(\tau_{B^{+}}/\tau_{B^{0}}) = 1.076 \pm 0.004$~\cite{PDG} is the ratio of the  lifetimes of the charged and neutral $B$ meson, and
$(f^{\pm}/f^{00}) = 1.053 \pm 0.052$~\cite{fpmf00}. The factor of 2 in Eq.~(\ref{eqn:ai}) arises because a $K^0$ forms a $K^0_{S}$ meson half the time. 
  Results are listed in Tables~\ref{tab:result} and  \ref{tab:result2}.

\begin{table}[!ht]
    \centering
     \caption{Reconstruction efficiency, signal yield, and branching fraction for each channel. Where two uncertainties are given, the first is statistical and the second is systematic, and the single uncertainty corresponds to the statistical component.}
    \begin{tabular}{l c c c c}
    \hline
    \hline
         Channel & $\epsilon$ (\%)&$n_{\text{\rm sig}}$ & & $\mathcal{B}$ ($10^{-5}$) \\
         \hline
        $B^{+} \to J/\psi(e^{+}e^{-}) K^{+} $         & 30.4&$3706 \pm 62$&&$6.00 \pm 0.10 \pm 0.19$\\
        $B^{+} \to J/\psi(\mu^{+}\mu^{-}) K^{+} $     &37.2&$4578 \pm 62$&&$6.06 \pm 0.09 \pm 0.19$ \\
        $B^{0} \to J/\psi(e^{+}e^{-}) K_{S}^{0} $     &20.4&$1052 \pm 33$&&$2.67 \pm 0.08 \pm 0.12$ \\
        $B^{0} \to J/\psi(\mu^{+}\mu^{-}) K_{S}^{0} $ &25.0&$1343 \pm 37$&&$2.78 \pm 0.08 \pm 0.12$ \\
     \hline
    \hline
    \end{tabular}
    \label{tab:result}
\end{table}

\begin{table}[!ht]
    \centering
        \caption{Measured $A_{I}$ and $R_{K}\left(J/\psi\right)$, where the first quoted uncertainty is statistical and the second is systematic.}
    \begin{tabular}{l c}
    \hline
    \hline
    Observable & Measured value  \\
    \hline
    $A_{I}\left(J/\psi (ee)K\right)$ & $-0.022 \pm 0.016 \pm 0.030$\\
    $A_{I} \left(J/\psi (\mu \mu)K \right)$ & $-0.006 \pm 0.015 \pm 0.030$ \\
    $R_{K^{+}}\left(J/\psi\right)$ &$\phantom{-}1.009 \pm 0.022 \pm 0.008$\\
    $R_{K_{S}^{0}} \left(J/\psi\right)  $ &$\phantom{-}1.042 \pm 0.042 \pm 0.008$\\
    \hline
    \hline
    \end{tabular}
    \label{tab:result2}
\end{table}

\section{Systematic Uncertainties}
\label{syst}

We consider several sources of systematic uncertainty contributing to the measurements. The nominal fits are performed with some PDF shape parameters fixed to values obtained from simulated events. To evaluate the systematic uncertainties associated with this, the data are refit by varying each fixed shape parameter by $\pm 1 \sigma$, with the obtained variation in signal yield taken as the systematic uncertainty.  Here $\sigma$ is the data driven uncertainty on each of these shape parameters, obtained by fitting to data. We assign 0.2\% uncertainty for the fixed shape parameters. An overall uncertainty of 1.5\% is assigned as a systematic uncertainty due to the estimate of the number of $B\overline{B}$ events. From the study of data--simulation differences in a sample of $e^+ e^- \to \tau^+ \tau^-$, we assign a  0.3\% systematic uncertainty for each charged track in the final state~\cite{tracking}. For the branching fraction measurements, we assign 0.9\% and 1.2\% systematic uncertainties for  $B^+$ and $B^0$ channels, respectively. For the measurement of $A_{I}$, since the leptons are common in both charged and neutral channels, their correction factor cancels, and we assign a 0.9\% systematic uncertainty for the charged kaon and two decay product tracks of the neutral kaon. The 0.1\% uncertainty in efficiency  due to the limited simulation sample size is also considered as a source of systematic uncertainty. 

From studies performed  with a sample of  $D^{*+} \to D^0 \left(\to K^{0}_{S} \pi^+ \pi^-\right) \pi^+$ decays, we observe that the data--simulation ratio of the $K^{0}_{S}$ reconstruction efficiency changes linearly as a function of the flight distance, which corresponds to an uncertainty of 0.4\% per cm of average flight length. The average flight length of $K^{0}_{S}$ for the signal channel is estimated to be 7.6\,cm using simulated signal events. Hence, we assign a 3.0\% systematic uncertainty for branching fraction and isospin asymmetry measurements involving a $K^{0}_{S}$ meson. A $D^{*+} \to  D^{0}\left(\to K^{-} \pi^{+}\right)\pi^{+}$ sample is used to estimate the difference in the kaon identification efficiency between data and simulation. The correction factors are calculated as functions of particle momentum and polar angle. We assign a 0.2\% systematic uncertainty  for measurements other than $R_{K}\left(J/\psi\right)$.  An inclusive $J/\psi \to \ell^{+} \ell^{-}$ sample is used to compute the difference in lepton identification efficiency between data and simulation, similar to the case of kaon identification. We assign a $0.6\%$  and $0.4\%$ systematic uncertainty for electrons and muons, respectively. 

As we reconstruct $K^0_S$ candidates from their $\pi^+\pi^-$ decays, we use  $\mathcal{B} \left(K^{0}_{S} \to \pi^{+}\pi^{-}\right) = (69.20 \pm 0.05)\% $  to account  for the other $K^{0}_{S}$ decay channels~\cite{PDG}.  We assign a 0.1\%  systematic uncertainty from this source to the  measurement of $\mathcal{B}$ of neutral channels and $A_{I}$. We use the measured branching fraction~\cite{fpmf00} of the $\Upsilon(4S)$ decay to a charged (neutral) $B$ meson pair, or $f^{\pm}$ ($f^{00}$). The associated systematic uncertainty for either of them is 2.6\%. The ratio $f^{\pm}/f^{00}$ contributes a 5.2\% systematic uncertainty for the $A_{I}$ measurement.  The assumed value of $(\tau_{B^+}/\tau_{B^-})$ leads to a 0.4\% uncertainty in the $A_I$ measurement~\cite{PDG}. 

We summarize the sources of systematic uncertainties for $\mathcal{B}$, $R_{K}\left(J/\psi\right)$, and $A_{I}$ in Table~\ref{table:Systematic}. The individual sources of uncertainties are assumed to be independent, and the corresponding contributions are added in 
quadrature to obtain the total uncertainty. The uncertainties for branching fractions and $R_K$ are relative, while the uncertainty for $A_{I}$ is absolute.

\begin{table}[H]

    \caption{ Relative systematic uncertainties (\%) on $\mathcal{B} \left(B \to  J/\psi K \right)$, $R_{K}\left(J/\psi\right)$, and absolute uncertainty on $A_{I}\left(B \to  J/\psi K\right)$. }
	\label{table:Systematic}
	\begin{center}
		\begin{tabular}{l  c c c c c c c c}
		\hline
		\hline

        Source &\multicolumn{4}{c}{$\mathcal{B} \left(B \to  KJ/\psi \right)$}   &\multicolumn{2}{c}{$R_{K} $}& \multicolumn{2}{c}{$A_{I} $}\\
        \cline{2-9}
         &$K^{+}$&$K^{+}$ & $K_{S}^{0}$ & $K_{S}^{0}$&$K^{+}$&$K^{0}$&&\\
        &$e^+e^-$&$\mu^+\mu^-$&$e^+e^-$&$\mu^+\mu^-$&&&$e^+e^-$&$\mu^+\mu^-$\\

        \hline   
        Number of $B\overline{B}$ events   & 1.5 & 1.5 & 1.5 &1.5& -- & -- & -- & --  \\
        PDF shape &  0.2& 0.2 &0.2  &0.2  &0.2&0.2&0.1&0.1\\
        Electron identification  &   $0.6$& --&    $0.6$ &-- &   $0.6$&   $0.6$&-- &--\\
        Muon identification  &  --&  $0.4$&  -- & $0.4$ &  $0.4$&  $0.4$&-- &--\\
        Kaon identification  & 0.2 & 0.2 & -- & --&--&--&0.1&0.1 \\
        $K^{0}_{S}$ reconstruction  & -- & --   & 3.0 & 3.0 &--&--&1.5&1.5 \\
        Tracking efficiency&0.9 &0.9  &1.2  &1.2&--&--&0.4&0.4  \\
        Simulation sample size  & 0.1 &0.1  &0.1  &0.1 &0.1&0.1&0.1& 0.1\\
        $\Upsilon(4S)$ branching fraction &2.6&2.6&2.6&2.6&--&--&2.6&2.6\\
        $(\tau_{B^{+}}/\tau_{B^{0}})$&--&--&--&--&--&--&0.2&0.2\\
        \hline
        Total &  $3.2$ & $3.2$  & $4.4$ &$4.4$ &$0.8$&$0.8$&$3.0$&$3.0$ \\
        \hline
        \hline
		\end{tabular}
	\end{center}
\end{table}

\section{Summary}
\label{summary}
We report the measurements of $\mathcal{B} \left(B\to J/\psi K\right)$, $A_{I}$, and $R_{K}$ of $B\to J/\psi(\ell^+ \ell^-)K$ decays performed by the Belle II experiment. The results
\begin{eqnarray*}
\mathcal{B} \left( B^{+} \to J/\psi(e^{+} e^{-}) K^{+}\right) &=& (6.00 \pm 0.10 \pm 0.19) \times 10^{-5},\\
\mathcal{B} \left( B^{+} \to J/\psi(\mu^{+} \mu^{-}) K^{+}\right) &=& (6.06 \pm 0.09 \pm 0.19) \times 10^{-5},\\
\mathcal{B} \left( B^{0} \to J/\psi(e^{+} e^{-}) K_{S}^{0} \right) &=& (2.67 \pm 0.08 \pm 0.12) \times 10^{-5},\\
\mathcal{B} \left( B^{0} \to J/\psi(\mu^{+} \mu^{-}) K_{S}^{0} \right) &=& (2.78 \pm 0.08 \pm 0.12) \times 10^{-5},\\
A_{I} \left( B \to J/\psi(e^{+} e^{-}) K\right) &=& -0.022 \pm 0.016 \pm 0.030,\\
A_{I} \left( B \to J/\psi(\mu^{+} \mu^{-}) K\right) &=& -0.006 \pm 0.015 \pm 0.030,\\
R_{K^{+}}\left(J/\psi\right) &=& 1.009 \pm 0.022 \pm 0.008, \text{ and}\\
R_{K^{0}}\left(J/\psi\right) &=& 1.042 \pm 0.042 \pm 0.008,
\end{eqnarray*}
are consistent with the world average values~\cite{PDG}. Because the signal selection efficiency of electron channels is comparable to that of muon channels, we can have uncertainties in the $R_K$ result that are almost equally contributed by the two lepton flavors. The $B\to J/\psi K$ decays will serve as control channels for the $B \to K \ell^+ \ell^-$ study.

\section{Acknowledgement}
We thank the SuperKEKB group for the excellent operation of the
accelerator, the KEK cryogenics group for the efficient
operation of the solenoid,  and the KEK computer group for on-site computing support.

\end{document}